\documentclass{iaus}
\usepackage{graphicx}

\title[Surface Brightness Variation of the Contact Binary SW Lac] 
{Surface Brightness Variation of the Contact Binary SW Lac: Clues From Doppler Imaging}

\author[H.V. \c{S}enavc{\i}]   
{Hakan Volkan \c{S}enavc{\i}$^1$}

\affiliation{$^1$University of Ankara, Faculty of Science, Department of Astronomy and Space Sciences, TR-06100 Tando\u{g}an-Ankara, TURKEY \\ email: {\tt hvsenavci@ankara.edu.tr}}

\pubyear{2011}
\volume{282}  
\pagerange{119--126}
\jname{From Interacting Binaries to Exoplanets: Essential Modeling Tools}
\editors{A.C. Editor, B.D. Editor \& C.E. Editor, eds.}
\begin{document}

\maketitle

\begin{abstract}
In this study, we present the preliminary light curve analysis of the contact binary SW Lac, using B, V light curves of the system spanning 2 years (2009 - 2010). During the spot modeling process, we used the information coming from the Doppler maps of the system, which was performed using the high resolution and phase dependent spectra obtained at the 2.1 m Otto Struve Telescope of the McDonald Observatory, in 2009. The results showed that the spot modeling from the light curve analysis are in accordance with the Doppler maps, while the non-circular spot modeling technique is needed in order to obtain much better and reliable spot models.
\keywords{techniques: photometric, (stars:) binaries: eclipsing, stars: spots}
\end{abstract}

\section{Introduction}

The light variability of the short-period contact binary SW Lac (P ~ 0.$^d$32, V$_{max}$=8.$^m$91) is very well known and studied by several investigators since its discovery by \cite[Miss Ashall (1918)]{leavitt18}. The first photoelectric UBV light curves of the system were obtained by \cite[Brownlee (1956)]{brownlee56}, who also pointed out the light curve asymmetries from cycle to cycle. These asymmetries were confirmed and attributed to the existence of cool spot regions by several authors (see \cite[Albayrak et al. 2004]{albayrak04}, \cite[Alton \& Terrell 2006]{alton06} and references there in). The spectral studies of the system including spectral classification, mass ratio determination and UV/X-ray region spectral analysis were carried out by several investigators, who revealed that the system is a W-type contact binary showing chromospheric and coronal activity (see \cite[\c{S}enavc{\i} et al. 2011]{senavci11} and references there in, for details).


The aim of this study is to perform the light curve analysis with the spot modeling, using the 2009 and 2010 light curves of the system with the help of the information coming from the Doppler maps obtained by \cite[\c{S}enavc{\i} et al. (2011)]{senavci11}.

\section{Observations and Data Reduction}

The 2009 and 2010 BV band light curves of the contact binary SW Lac were obtained at the Ankara University Observatory, using an Apogee Alta U47 CCD camera attached to a 40 cm Schmidt-Cassegrain telescope. BD+37$^{\circ}$ 4715 and BD+37$^{\circ}$ 4711 were chosen as comparison and check stars, respectively. The nightly extinction coefficients for each passband were determined by using the observations of the comparison star. A total of 700 and 995 data points were obtained in each passband, while the probable error of a single observation point was estimated to be $\pm 0.003/0.004$ and $\pm 0.004/0.004$ for 2009 and 2010 BV bands, respectively. 

\section{The Light Curve Analysis}

\begin{figure}[h]
\begin{center}
 \includegraphics[width=2.6in]{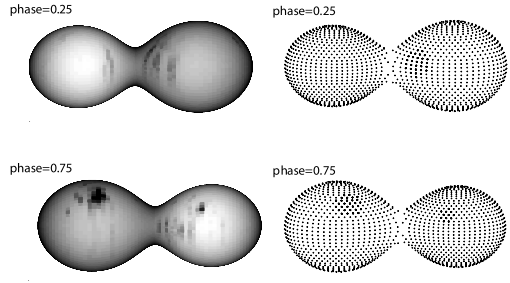} 
\end{center}
 \caption{The Doppler maps and the adopted spots for LC modeling of the system.}
   \label{fig1}
\end{figure}

\begin{figure}[h]
\hspace*{1.7 cm}
 \includegraphics[width=1.85in]{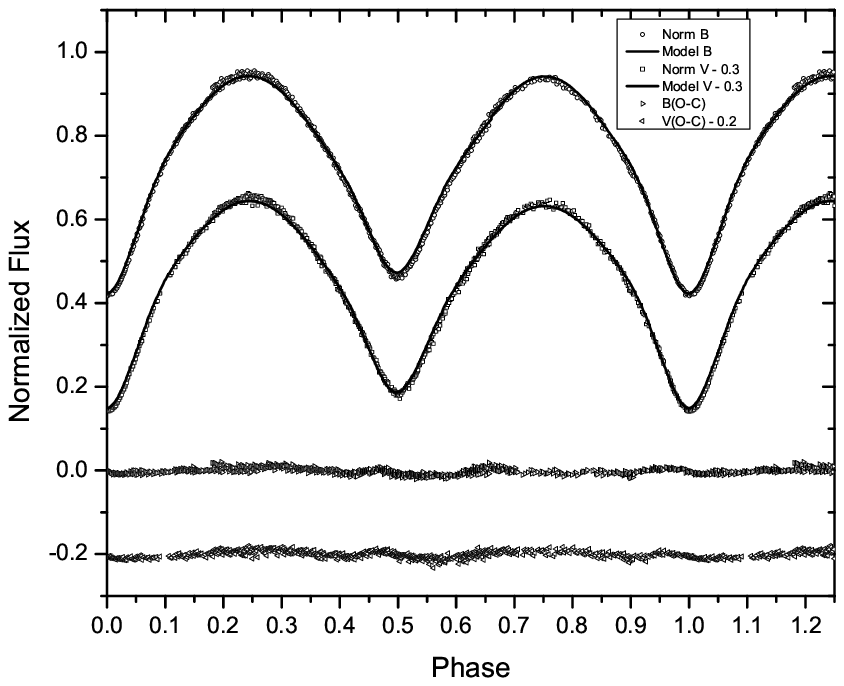} 
 \includegraphics[width=1.85in]{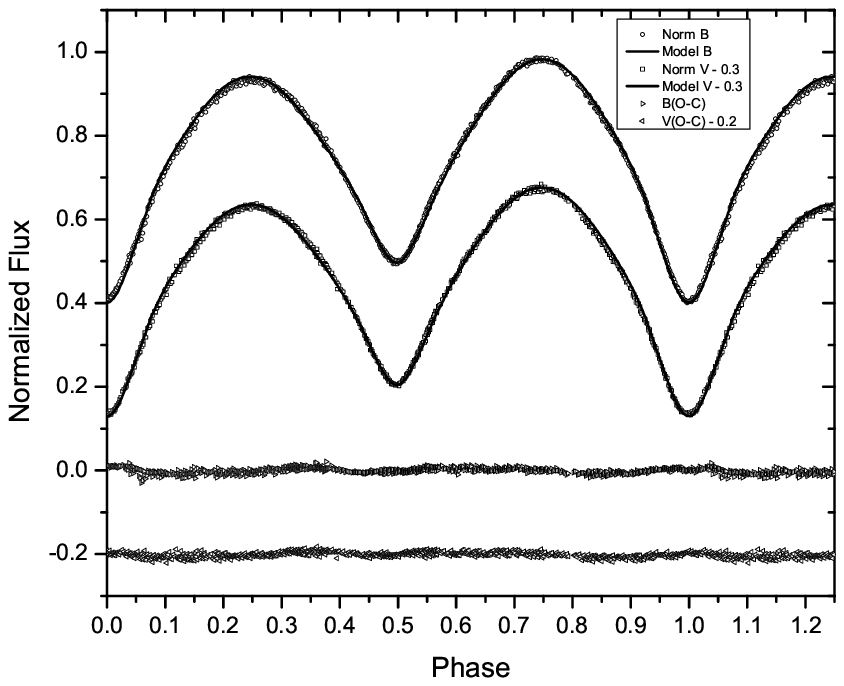}
 \caption{Observational and theoretical light curves with O-C residuals for 2009 and 2010.}
   \label{fig2}
\end{figure}

The 2009 and 2010 BV light curves were analysed simultaneously with the radial velocity curves of the system obtained by \cite[Rucinski et al. (2005)]{rucinski05} using the interface version of the Wilson-Devinney code (\cite[Wilson \& Devinney 1971]{wilson71}), PHOEBE (\cite[Prsa \& Zwitter 2005]{prsa05}). Since the surface reconstructions of the system were performed using the time series spectra obtained in 2009, we first adopted the spot modeling, as three main circular spot regions, to 2009 light curves and carried out the LC modeling (see Fig.\ref{fig1}). The results from the LC and spot modeling were represented in Fig.\ref{fig2}.

\section{Conclusion}

The analysis showed that the theoretical light curves are compatible with the observed ones, though the circular spot modeling was performed. However, in order to perform more reliable spot modeling, a code with a none circular shaped spot approximation is needed as the Doppler maps clearly show us the spots are not circular.


\end{document}